# Cross-Frequency Coupling During Thermoacoustic Oscillations in a Pressurized Aeronautical Gas Turbine Model Combustor

Mitchell L. Passarelli [a,*], Timothy M. Wabel [a], Arin Cross [b], Krishna Venkatesan [b], Adam M. Steinberg [c]

[a] *Institute for Aerospace Studies, University of Toronto, 4925 Dufferin St, Toronto, ON M3H 5T6, Canada*
[b] *GE Research, 1 Research Cir, Niskayuna, NY 12309, USA*
[c] *School of Aerospace Engineering, Georgia Institute of Technology, 270 Ferst Dr, Atlanta, GA 30332, USA*

Submitted to the Gas Turbine and Rocket Engine Combustion Colloquium

| Section | Main Text | Equations | References | Figures and Captions Total | Total |
|---|---|---|---|---|---|
| Word Count | 3,399 | 76 | 576.84 | 1,701.18 | 5,753.02 |

Figure Word Equivalents (including captions):
- Fig. 1 = 116.84
- Fig. 2 = 197.76
- Fig. 3 = 242.88
- Fig. 4 = 262.2
- Fig. 5 = 221.96
- Fig. 6 = 398.38
- Fig. 7 = 261.16

[*] Corresponding author.
 *E-mail address:* mpassarelli3@gatech.edu (M. Passarelli).

# Cross-Frequency Coupling During Thermoacoustic Oscillations in a Pressurized Aeronautical Gas Turbine Model Combustor

Mitchell L. Passarelli [a,†], Timothy M. Wabel [a], Arin Cross [b], Krishna Venkatesan [b], Adam M. Steinberg [c]

[a] *Institute for Aerospace Studies, University of Toronto, 4925 Dufferin St, Toronto, ON M3H 5T6, Canada*
[b] *GE Research, 1 Research Cir, Niskayuna, NY 12309, USA*
[c] *School of Aerospace Engineering, Georgia Institute of Technology, 270 Ferst Dr, Atlanta, GA 30332, USA*

**Abstract**

This paper demonstrates cross-frequency coupling between pressure, heat release rate, fuel spray and velocity oscillations in a model aeronautical gas turbine combustor operating at a pressure of approximately 10 atm. Heat release rate was characterized by 10 kHz chemiluminescence (CL) imaging of several species. Stereoscopic particle image velocimetry and laser Mie scattering from the fuel droplets were used to measure the gas velocity and spray dynamics, respectively, at 5 kHz. The pressure fluctuations were dominated by oscillations at a frequency $f_0$, whereas the spray, CL and velocity oscillated at approximately $2f_0$. All of these oscillations were nonstationary, exhibiting changes in frequency and amplitude. Comparing the time evolution of the dominant frequencies and amplitudes indicates a behavior consistent with mutually coupled self-oscillators; the observed dynamics of the 1:2 super-harmonic coupling is consistent synchronization via oscillation death. Furthermore, increases in the frequency of the ca. $f_0$ velocity oscillations away from the harmonic ratio (increased frequency detuning) were correlated with decreases in the power of the $f_0$ pressure oscillations. The corresponding nonreacting flow had a natural hydrodynamic mode at a frequency slightly greater than $2f_0$. Hence, the data are consistent with the $f_0$ acoustic mode 'pulling' the hydrodynamic frequency towards the super-harmonic ratio.

---

[†] Corresponding author.
 *E-mail address:* mpassarelli3@gatech.edu (M. Passarelli).

# 1. Introduction

Prediction and mitigation of thermoacoustic instabilities remains a major challenge in the design of low-emission gas turbine combustors. Linear thermoacoustic models, generally employing flame transfer functions, have proven useful for predicting instability frequencies, pressure mode shapes and linear growth rates [1-3]. However, nonlinear methods are necessary to predict amplitudes and frequencies at the limit-cycle. Nonlinearity in the flame response to perturbations may be captured by flame describing functions (FDFs), which sensitize the modelled response of heat release fluctuations to the perturbation amplitude [4-6]. In such methods, it is assumed that the flame responds only at the frequency of the (linear) acoustic forcing; sub- and super-harmonics generated by the flame's nonlinear response are not considered. However, little direct evaluation of this assumption has been performed [5]. This paper presents experimental evidence of the importance of nonlinear harmonic responses in an aeronautical gas turbine combustor operating at practically-relevant conditions.

Illingworth and Juniper [5] performed an analytical study of super-harmonic interactions using a simple nonlinear flame model, coupled to duct acoustics. While this model indicated that FDFs could perform well even when contributions from harmonics were significant, the model did not consider the even harmonics; excitation at a frequency $f_0$ could excite $3f_0$ but not $2f_0$, etc. Kim [7] experimentally studied interactions between the fundamental frequency and higher harmonics in atmospheric pressure premixed and partially-premixed swirl flames. They found that higher harmonics contain information that is critical for understanding the mechanisms driving the limit-cycle oscillations, particularly for their higher amplitude cases, and concluded that FDFs may face challenges in such situations. Haeringer et al. [8] recently proposed an extended FDF approach that relates higher harmonics of the heat release rate to the forcing frequency and demonstrated this method for a laminar premixed flame. Nevertheless, the manifestation and importance of harmonic interactions remains unclear, particularly for realistic gas turbine combustors.

The analysis of cross-frequency coupling—of which harmonics are an example—can be accomplished using synchronization theory, which previously has been used to analyze various aspects of combustion dynamics [9-13]. However, these studies focused on same-frequency, also called 1:1, coupling. For example, Mondal et al. [9] studied the transition from aperiodic noise to constant amplitude limit cycle oscillations with changing conditions, which occurred through a range of conditions exhibiting intermittent bursts of coherent oscillations. Limit cycle oscillations corresponded to perfect phase synchronization, whereas intermittent oscillations corresponded to intermittent synchronization between the pressure and heat release rate.

More generally, systems influenced by cross-frequency, or $m:n$, coupling exhibit qualitatively similar characteristics and behaviours as seen during 1:1 coupling [14, 15]. Such nonlinear cross-frequency coupling generally is associated with transient (intermittent) oscillations and has been studied across many fields [16-18]. We note that $m:n$ coupling can fundamentally occur for any integer values of $m$ and $n$; they need not be direct sub- or super-harmonics [14, 17, 19]. However, as will be discussed below, sub-harmonics ($m$: 1) and super-harmonics (1: $n$) tend to most readily couple. Moreover, synchronization between oscillators is a sufficient, but not necessary, attribute to demonstrate that the oscillators are coupled; it is the cross-frequency coupling that permits synchronization [14].

While numerous experimental studies have characterized 1:1 limit-cycle self-excited thermoacoustic oscillations, the majority of these have been performed at atmospheric pressure with gaseous fuels to facilitate the use of laser diagnostic techniques [20, 21]. Practical gas turbine combustors operate at elevated pressures and, in the case of aeronautical engines, with liquid fuels. Such engines have much higher power densities than lab-scale atmospheric pressure burners, which may increase the significance of nonlinear cross-frequency coupling. Recent studies [22-26] have demonstrated the successful application of high-speed optical diagnostics to study combustion dynamics at elevated pressures, though cross-frequency coupling has not been analyzed.

This paper demonstrates the effects of cross-frequency coupling between the pressure, chemiluminescence, fuel spray and gas velocity in a model aeronautical gas turbine combustor operating at practically-relevant conditions. In particular, the intermittent fluctuations observed in the pressure are related to coupling with the other quantities, which oscillate at the first super-harmonic of the pressure.

## 2. Experimental Setup

The experiments were performed at the GE Research on an optically accessible model gas turbine combustor, a schematic of which is shown in Fig. 1. The combustor operates at fuel-rich conditions using liquid Jet A fuel supplied via a swirl nozzle. Experiments were performed at air preheat temperatures of approximately 500 K and pressures around 1 MPa. Different thermoacoustic behaviours were achieved at different fuel flow rates; the data analyzed here is from a case with a thermal power of approximately 0.4 MW.

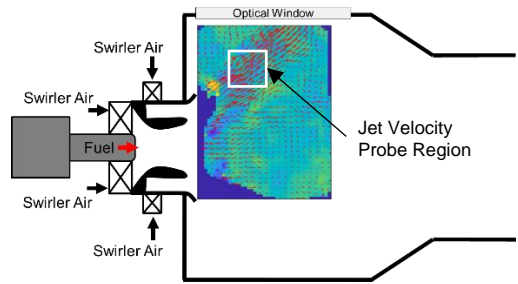

Fig. 1. Schematic in profile of test rig with sample instantaneous velocity field.

High-speed chemiluminescence (CL) imaging in four different spectral ranges, sampled at 10 kHz, was used to characterize the heat release rate integrated along the line-of-sight. Specifically, OH* at 308 ± 5 nm and CH* at 430 ± 5 nm were measured as metrics of the heat release rate, while $C_2$* at 470 ± 5 nm and chemiluminescence from larger species (e.g. $CO_2$*) at 341 ± 5 and 403 ± 5 nm were measured to assess the background spectrum [27-29]. Two Photron SA-Z cameras enabled simultaneous imaging of two species at a time with narrow bandpass filters to isolate the desired signals. The cameras were equipped with commercial objective lenses (Nikkor 105 mm UV, f/# = 16) and lens-coupled image intensifiers (Invisible Vision UVi 1850-10 and Video Scope International VS4-1845HS, gate time of 50 µs). The $C_2$* and large-species measurements were obtained primarily to assess whether the observed intensities in the CH* and OH* bands correspond to CH* and OH* emissions or broadband emissions due to the rich operating conditions. It was found that instantaneous and mean images from the large-body, $C_2$* and CH* bands exhibited similar structures, indicating that most of the signal measured in the 'CH* band' is actually from broadband sources. This was confirmed based on wide-spectrum measurements of the spatially- and temporally-integrated combustor emissions made using a portable spectrometer. As such, the OH* signal was selected as the most appropriate metric of the heat release rate oscillations, which is consistent with previous results [22]. It is noted that, given the non-premixed and highly turbulent nature of the flame, OH* chemiluminescence can only be used as a qualitative indicator of the location and dynamics of the heat release rate.

Stereoscopic particle image velocimetry (S-PIV) was used to characterize the gas-phase velocity in the combustion chamber at a rate of 5 kHz. Details of the system are provided in Ref. [22] and are only summarized here. The flow was illuminated using the double-pulsed output of a second harmonic Nd:YAG laser (Quantronix Hawk-Duo), with a pulse separation time of 5 µs. A three cylindrical-lens telescope formed the beam into a sheet with a thickness of approximately 2 mm at the beam-waist in the test section. The imaging system consisted of two high-speed cameras (Photron SA-5) equipped with 532 ± 2.5 nm bandpass filters, objective lenses (Tamron 180 mm, f/# = 5.6), and Scheimpflug adapters (LaVision). Scattered light from ca. 5 µm $ZrO_2$ particles and the fuel droplets was collected into the cameras for measurements of the gas phase velocity and spray dynamics, respectively.

The fuel entered the combustor with droplet sizes much larger than the PIV seed particles, and hence scattered more light. As described in Kheirkhah et al. [22], an image segmentation method was used to separate the flow tracer particles from the droplets. While this method is unable to distinguish droplets that are similar in size to the PIV seed, such droplets are expected to follow the flow as well as the seed and therefore were included in the PIV particle images. It is noted that the Stokes number of PIV seed relative to the Kolmogorov time scale was too large for the seed to accurately follow the fine-scale velocity fluctuations. However, the seed was sufficient to follow the larger-scale thermoacoustically-coupled oscillations of interest here.

S-PIV processing was performed using commercial software (LaVision DaVis 8.4). Raw particles images were imported, segmented to isolate seed particles, and then spatially and temporally filtered to remove background and flame luminosity. Camera calibration was performed using a 3D dot-target and refined using a particle image-based self-calibration routine. A multi-pass vector processing algorithm was used, with adaptive window shape and size and an iterative vector filter. For the final pass, the interrogation boxes were 48×48 pixels and overlapped by 75%; the resolution and vector spacing were 3.73 mm and 0.93 mm, respectively. Spurious vectors were removed using a median filter outlier detection algorithm. Remaining gaps in the vector fields were filled using gappy POD (GPOD) [30], excluding regions where less than 50% of the snapshots had vectors.

## 3. Analysis Techniques

### 3.1. *Dynamics of Coupled Self-Oscillators*

To analyze cross-frequency coupling, we introduce the harmonic ratio $m:n$ between the oscillation frequencies, defined as [14, 15]

$$\frac{f_1}{f_2} \approx \frac{m}{n}, \qquad m, n \in \mathbb{N} \tag{1}$$

The harmonic ratio must be accounted for when computing parameters such as the phase difference, $\Delta\phi$, and frequency detuning, $\Delta f$:

$$\Delta\phi = n\phi_1 - m\phi_2 \tag{2}$$

$$\Delta f = mf_2 - nf_1 \tag{3}$$

It should be noted that higher-order (cross-frequency) coupling (when at least one of $m$ or $n$ is not 1) is enabled by the fact that the coupled self-oscillators are not perfectly harmonic (sinusoidal) oscillators, by definition [14].

Two parameters characterizing mutually coupled oscillators are the dissipative and reactive coupling strengths, $B_D$ and $B_R$ respectively. Dissipative coupling acts to synchronize self-oscillators by damping the difference in 'velocities', while reactive coupling works by modulating the 'positions' of the oscillators. Together with $\Delta f$, the coupling strengths define the regions of synchronization for each harmonic ratio. These regions are generally V-shaped, revealing an increasing tolerance to $\Delta f$ with increasing coupling strengths, illustrated by the schematic in Fig. 2. Indicated in Fig. 2 are the regions corresponding to unsynchronized and synchronized dynamics. The suppression region is where suppression of the natural dynamics occurs and corresponds to complete synchronization. Crossing the borders between the phase-locking and suppression regions causes a bifurcation in the dynamics of the system.

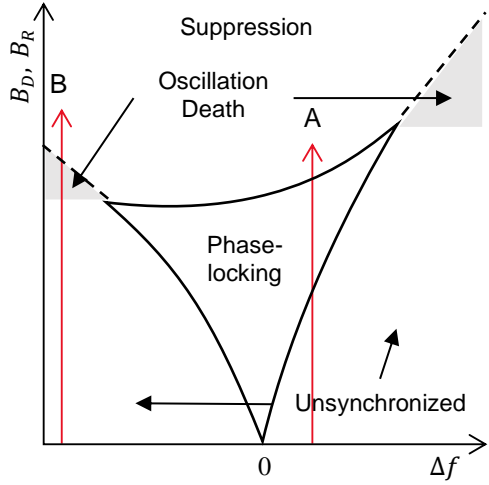

Fig. 2. Schematic representation of $m:n$ synchronization region. Arrow A denotes the phase-locking route; B denotes the oscillation death route. Solid lines denote saddle-node bifurcations and dashed lines indicate torus-birth bifurcations.

Two basic routes to synchronization are indicated by the vertical arrows in in Fig. 2, *viz.* phase-locking (A) and oscillation death (B). Along the phase-locking route, increasing $B_D$ causes the frequencies of both oscillators to be drawn towards each other until they coincide, yielding phase-locked dynamics. Increasing $B_R$ instead causes both frequencies to increase, with the slower oscillator speeding up until it 'catches' the faster one. Once the frequencies of the oscillators coincide (i.e., $\Delta f = 0$), the oscillators are synchronized. For both types of coupling, the power of the oscillations at the coupled frequencies remain relatively constant.

Along the oscillation death route (B in Fig. 2), an increase in $B_D$ or $B_R$ is associated with a decrease in the amplitudes of the oscillations—with frequencies shifting much less than along the phase-locking route—until both oscillators stop. Further increasing the coupling strengths will eventually restart the oscillations but at a common, synchronized frequency. The direction of movement in the frequencies before oscillation death occurs is consistent with that observed along the phase-locking route for both coupling types. Interestingly, the ratio between the synchronized frequencies for purely reactive coupling ($B_D = 0$) is slightly higher than the ratio for the nearest harmonic ratio for both routes [14]. The behaviours observed along each route to synchronization are universal and applicable to any self-oscillating system [14].

Unfortunately, it is usually not possible to measure $B_D$ and $B_R$ experimentally [15]. One must instead inspect the power spectra, phase portraits, phase space reconstructions and the oscillations themselves to qualify $B_D$ and $B_R$ in a given experiment [14]. Fortunately, as described above, distinct effects on the system should be observed for increasing $B_D$ or $B_R$ at a given $\Delta f$. The only complication with analyzing real systems is that there will be a combination of both reactive and dissipative coupling. Additionally, these characteristic behaviours are indicative of cross-frequency interactions even if synchronization is not observed.

*3.2. Spectral Proper Orthogonal Decomposition*

To perform the analysis described above, it is useful to identify and isolate the dynamical aspects of each measured signal that oscillate in the frequency ranges of interest. Here, spectral proper orthogonal decomposition (SPOD) [31] is used for this purpose. The SPOD algorithm is identical to that of classical POD, with the additional step of lowpass filtering along the diagonals of the correlation matrix before performing the eigenvalue decomposition. The filtering smooths temporal fluctuations in the frequency of each mode, acting like a bandpass filter to eliminate the mode blending exhibited by POD and distribute the energy contained in those fluctuations to the other modes [31]. As a result, the SPOD modes oscillate within distinct frequency ranges and, unlike a bandpass filter or phase-averaging, no information is lost in the SPOD operation. SPOD was found to be more appropriate than dynamic mode decomposition (DMD) due to the non-stationary nature of the oscillations (see Section 4).

Following Sieber et al. [31], a Gaussian filter was used along with their mode pairing algorithm to extract the dominant dynamics based on their spectral coherence. The filter width was tuned for CL, spray, pressure and gas velocity individually.

## 4. Results & Discussion

To begin, it is necessary to articulate the spectral content of the various signals. The temporal dynamics of the pressure and gas velocity fluctuations are shown in the spectrograms of Fig. 3, with frequencies normalized by the dominant acoustic frequency $f_0$. The gas velocity spectrogram is computed at the location in the inflowing air jet indicated in Fig. 1, though similar spectral content was found at all locations. The spectrograms were computed using Welch's method with a temporally sliding window, where each window overlapped with its neighbours by 75% and the window size was based on the sampling frequencies of the measurements. The window size and overlap were chosen to balance the frequency and time resolutions of the spectrograms. The features in the resulting plots are independent of the chosen parameters over a reasonable range.

The temporally resolved spectral content of the OH* and spray dynamics is difficult to visualize using spectrograms due to the relatively low amplitude of the coherent oscillations compared to the turbulent fluctuations. Instead, the overall spectral content of these signals is shown in the SPOD spectra in Fig. 4, where each dot represents a mode-pair and the size and colour of the dot represents the magnitude of the spectral coherence of that mode-pair. As such, the brightest and largest circles denote the most coherent structures in the data.

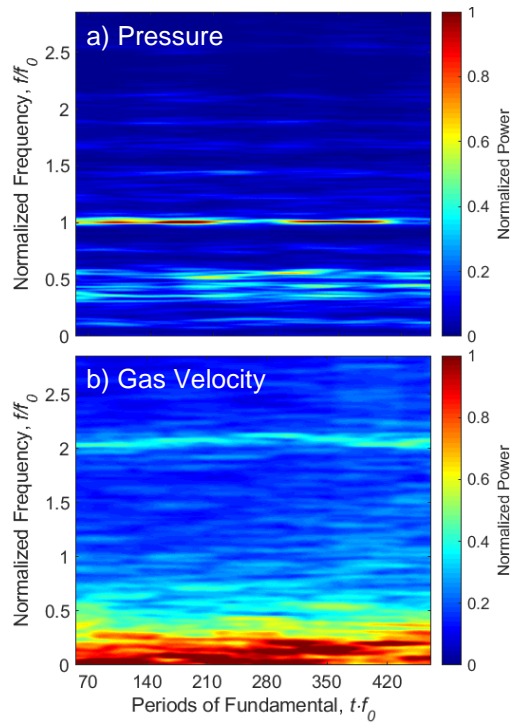

Fig. 3. Spectrograms of a) pressure and b) gas velocity fluctuations.

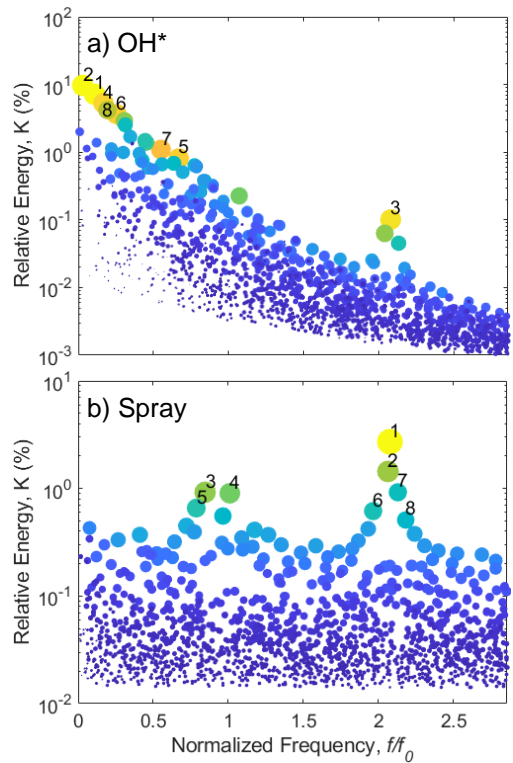

Fig. 4. SPOD spectra of a) OH* CL and b) fuel spray with eight most coherent modes indicated.

These figures clearly demonstrate the importance of harmonic interactions in this system. Whereas the dominant pressure oscillations are at $f_0$, there are no coherent velocity oscillations at this frequency; the velocity oscillations are at approximately $2f_0$. Similarly, the OH* CL exhibited more coherence and higher energy relative to the turbulent fluctuations at $2f_0$ than at $f_0$.

It is noted that the mode shape corresponding to the mode pair labelled '3' in Fig. 4a (not shown) has a similar structure to the SPOD velocity mode at $2f_0$ (described below). The spray dynamics (Fig. 4b) also exhibit coherent dynamics at $2f_0$, which have more energy relative to the background than the spray dynamics at $f_0$.

Figure 3 also shows that there is significant drift in the frequency of the velocity oscillations near $2f_0$, and that the timing of this drift coincides with changes in the power of the pressure oscillations at $f_0$. This suggests that the $2f_0$ velocity oscillations are coupled with the $f_0$ pressure oscillations.

To better visualize this coupled behaviour, Fig. 5 shows the temporal evolution of the frequencies and amplitudes of the various quantities at the frequencies of interest. These attributes were extracted from the spectrograms by finding the peak of the power ridge in the frequency band around $f_0$ for pressure and $2f_0$ for velocity, OH* and spray. Note that the sudden steps in frequency occur due to features such as that occurring around $tf_0 \approx 280$ in Fig. 3a.

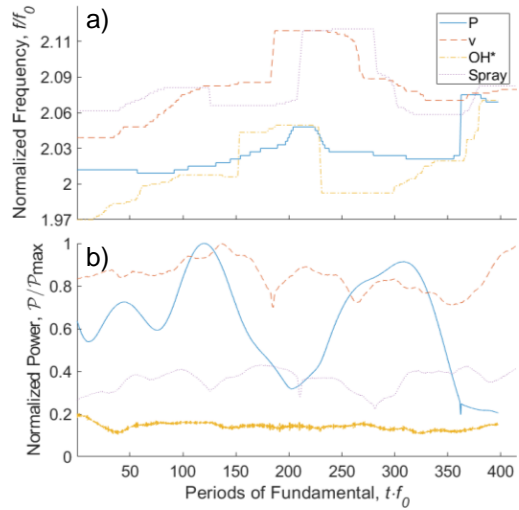

Fig. 5. Temporal evolution of a) peak frequencies and b) power $\mathcal{P}$ of peak frequencies (normalized by maximum power for all time in corresponding spectrogram $\mathcal{P}_{max}$). Pressure frequency is doubled to account for 1:2 harmonic ratio and OH* power is multiplied 20 times. 'P' represents pressure data and 'v' represents the velocity data.

Over periods 75-125 in Fig. 5, the $f_0$ pressure oscillations increase in power and all the frequencies increase slightly. Different trends are observed in the following intervals. Over periods 125-200, the frequencies increase while the pressure power decreases, and the opposite occurs over periods 200-300. Note that the frequency increase in the velocity, OH* and spray oscillations over $tf_0 = 75\text{-}200$ is much larger than that for the pressure oscillations. The behaviours in both intervals are consistent with reactive coupling along the route to oscillation death, i.e., when $\Delta f$ (given by the difference in the lines) between the pressure and velocity is relatively large [14].

Along the oscillation death route, the power of the oscillations tends to zero as the coupling strength increases. As the frequency detuning changes, however, it is possible for other peaks to appear in the power spectra, especially for higher-order coupling [14]. Comparing Figs. 3 and 5 shows that other peaks do appear and their appearance is responsible for the sudden frequency jumps in Fig. 5a. Thus, these jumps likely indicate that the oscillations are close to the oscillation death boundary.

This interpretation is further supported by the power spectra shown in Fig. 6. Each spectrum represents a vertical slice (along the frequency axis) at a particular time in the spectrogram for a given quantity. The times for each spectrum correspond to before approaching oscillation death ($tf_0 = 100$, Fig. 6a), near oscillation death ($tf_0 = 200$, Fig. 6b) and after oscillation death ($tf_0 = 300$, Fig. 6c).

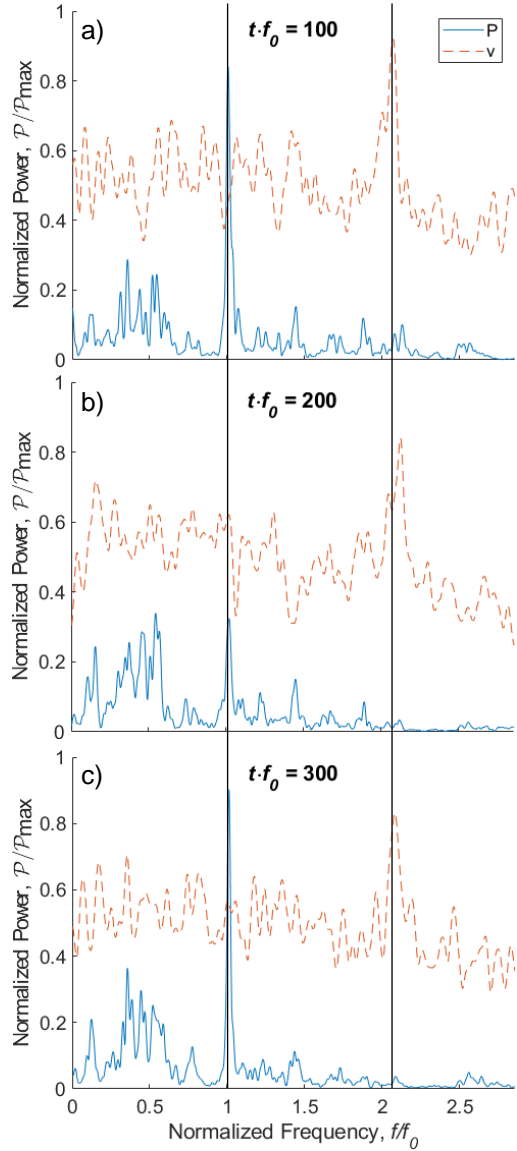

Fig. 6. Temporal evolution of power spectra (power $\mathcal{P}$ normalized by maximum power for all time in corresponding spectrogram $\mathcal{P}_{max}$) for times a) before nearing, b) near to and c) after nearing oscillation death. Vertical lines indicate the peak frequencies.

Between $tf_0 = 100$ and 200, the powers of the pressure oscillations at $f_0$ and velocity oscillations at $2f_0$ decrease, albeit much less so for the velocity. As the oscillations proceed and the amplitude of the $f_0$ pressure fluctuations increase, the velocity frequency near, but slightly higher than, $2f_0$ is pulled back towards $2f_0$ (compare Fig. 6b and c). This evolution is mostly consistent with expectations for increasing the reactive coupling strength between two self-oscillators along the oscillation death route [14].

However, unlike theoretical predictions, the power of the velocity oscillations at $2f_0$ varies only slightly as the system nears oscillation death. One possible explanation for this discrepancy is that the coupling effect of the pressure on the velocity is weaker than the effect of the velocity on the pressure. Such a scenario could arise if the oscillations in the velocity are due to a natural hydrodynamic instability and are thus driven by independently from the pressure and heat release rate dynamics. An SPOD analysis of a nonreacting data set reveals the same mode near $2f_0$ in the velocity, but with lower amplitude and higher frequency. The SPOD spectrum of the velocity for the nonreacting data set is shown in Fig. 7a, where mode pair 1 represents the nonreacting mode at around $2f_0$. The reacting and nonreacting SPOD mode shapes are shown in Fig. 7b and c, respectively. Both sets of mode pairs represent the same physical structure in both data sets, which is a double-helix vortical mode.

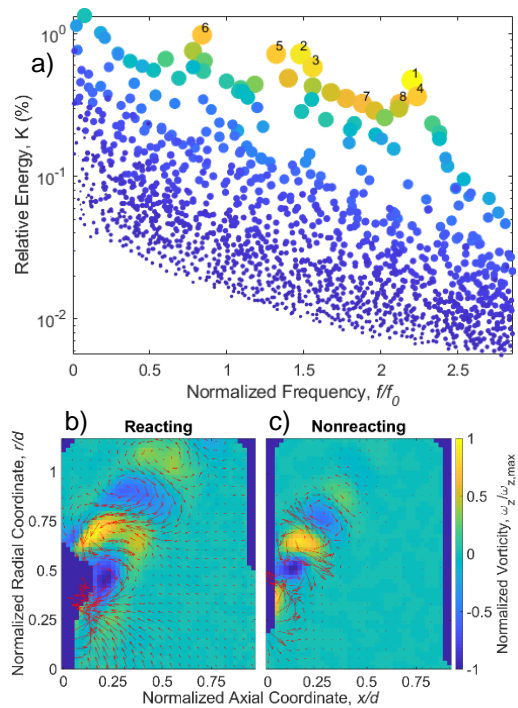

Fig. 7. a) Gas velocity SPOD spectrum from nonreacting data set with eight most coherent modes indicated. SPOD mode shapes for b) reacting and c) nonreacting gas velocity, where $d$ is the nozzle flare diameter.

These results indicate that the velocity field has a natural oscillatory mode at a frequency that is slightly greater than twice the excited acoustic mode in the combustor. Cross-frequency coupling with the pressure oscillations causes these frequencies to shift towards each other (with respect to the 1:2 harmonic ratio), with the frequency detuning depending on the coupling strength. This is also consistent with the observed behaviours of the OH* and spray oscillations. Hence, the dynamics of this combustor are strongly influenced by cross-frequency interactions acting along the oscillation death route, causing both the frequencies to shift relative to each other and the power of the oscillations to change as the coupling strength changes.

## 5. Conclusion

This paper investigates the effects of cross-frequency coupling between pressure, heat release rate, fuel spray and gas velocity fluctuations using synchronization theory as a framework. The data was collected on a model aeronautical gas turbine combustor operating at flight-relevant conditions. The data contain distinct pressure oscillations at frequency $f_0$ while the velocity, heat release rate (as indicated by OH* CL) and fuel spray oscillate near $2f_0$. The temporal evolutions of the frequencies and powers of the dynamics at $f_0$ and $2f_0$ are characteristic of mutually coupled self-oscillators close to the oscillation death boundary with primarily reactive coupling.

In general, these results show the importance of accounting-for and modelling nonlinear mechanisms in high power density systems. Cross-frequency coupling is one such nonlinear mechanism that produces rich oscillatory dynamics. This paper reports that cross-frequency interactions in thermoacoustic instabilities can manifest as amplitude modulations or frequency shifting. Future work should attempt to measure the coupling strength and to map out the regions of synchronization for higher-order coupling as presented in this study.

**Acknowledgements**

This work was supported by NSERC Canada under grant CRDPJ 515554-17 and by GE.

**List of Captions**